\documentclass[twocolumn,showpacs,preprintnumbers]{revtex4}
\usepackage{epsfig}
\usepackage{bm}

\newcommand{\beq}{\begin{equation}}
\newcommand{\eeq}{\end{equation}}

\newcommand{\etal}{{\rm et al.}}

\newcommand{\dams}{\Delta a_\mu({\rm SUSY})}

\newcommand{\app}[3]{Astropart.\ Phys.\ {\bf #1}, #3 (#2)}

\newcommand{\hepph}[1]{{\tt hep-ph/#1}}
\newcommand{\astroph}[1]{{\tt astro-ph/#1}}
\newcommand{\prep}[3]{Phys.\ Rep.\ {\bf #1}, #3 (#2)}
\newcommand{\plb}[3]{Phys.\ Lett.\ B\ {\bf #1}, #3 (#2)}
\newcommand{\npb}[3]{Nucl.\ Phys.\ B\ {\bf #1}, #3 (#2)}
\newcommand{\cpc}[3]{Comm.\ Phys.\ Comm.\ {\bf #1}, #3 (#2)}
\renewcommand{\apj}[3]{Astrophys.\ J.\ {\bf #1}, #3 (#2)}

\renewcommand{\prl}[3]{Phys.\ Rev.\ Lett. {\bf #1}, #3 (#2)}
\renewcommand{\prd}[3]{Phys.\ Rev.\ D\ {\bf #1}, #3 (#2)}
\renewcommand{\rmp}[3]{Rev.\ Mod.\ Phys.\ {\bf #1}, #3 (#2)}

\begin{document}


\title{Improved constraints on supersymmetric dark matter from muon g-2}

\author{E.~A.~Baltz}
\affiliation{ISCAP, Columbia Astrophysics Laboratory, 550 W 120th St., Mail
Code 5247, New York, NY 10027}
\email{eabaltz@physics.columbia.edu}
\author{P.~Gondolo} \affiliation{Department of Physics, Case Western Reserve
University, 10900 Euclid Ave., Cleveland, OH 44106-7079}
\email{pxg26@po.cwru.edu}

\date{\today}

\begin{abstract}
The new measurement of the anomalous magnetic moment of the muon by the
Brookhaven AGS experiment 821 again shows a discrepancy with the Standard Model
value.  We investigate the consequences of these new data for neutralino dark
matter, updating and extending our previous work [E.~A.~Baltz and P.~Gondolo,
Phys.~Rev.~Lett.~{\bf 86}, 5004 (2001)].  The measurement excludes the Standard
Model value at $2.6\sigma$ confidence.  Taking the discrepancy as a sign of
supersymmetry, we find that the lightest superpartner must be relatively light
and it must have a relatively high elastic scattering cross section with
nucleons, which brings it almost within reach of proposed direct dark matter
searches.  The SUSY signal from neutrino telescopes correlates fairly well with
the elastic scattering cross section.  The rate of cosmic ray antideuterons
tends to be large in the allowed models, but the constraint has little effect
on the rate of gamma ray lines.  We stress that being more conservative may
eliminate the discrepancy, but it does not eliminate the possibility of high
astrophysical detection rates.
\end{abstract}


\maketitle

\section{introduction}

In early 2001, the Brookhaven AGS experiment 821 measured the anomalous
magnetic moment of the muon $a_\mu=(g-2)/2$ with three times higher accuracy
than it was previously known \cite{DATA}.  Their result disagreed with the
Standard Model prediction at greater than 2.6$\sigma$.  However, a sign error
in the calculation of the hadronic light-by-light contribution to $a_\mu$ was
discovered, reducing the discrepancy to 1.6$\sigma$ \cite{LBLerror}.  Recently,
the same collaboration has released a result with much improved statistics
\cite{newdata}, and there is again a discrepancy at the $2.6\sigma$ level.
Supersymmetric particles can give significant corrections to $a_\mu$
\cite{a_mu_old,moroi,SM}, thus the Brookhaven measurement is an important
constraint on supersymmetric models.  There has been a substantial literature
on this topic since the announcement of the discrepancy
\cite{bg2001,otherpapers}, discussing various consequences of the older
measurement.  In this paper, we update the results of \cite{bg2001} concerning
the implications of the Brookhaven data for supersymmetric cold dark matter,
assuming that supersymmetry is the only relevant physics outside of the
Standard Model.

There are two significant assumptions in our discussion.  The first is that the
Standard Model prediction for the muon anomalous magnetic moment is somewhat
disputed, primarily in the hadronic contribution.  This was clearly
demonstrated in the sign error discovered in the last year.  The hadronic error
is a very significant part of the error budget when comparing the Brookhaven
results to the Standard Model.  In fact it has been claimed that the Standard
Model errors have been significantly underestimated \cite{nodiscrepancy}, but
this claim has been refuted \cite{newSM}.  Furthermore, there are new
evaluations of the hadronic vacuum polarization from firstly
$e^+e^-\rightarrow{\rm hadrons}$ indicating a larger discrepancy $(3.6\sigma)$
and secondly hadronic tau lepton decays indicating a smaller discrepancy
$(1.3\sigma)$ \cite{newdata}.  The second caveat is that supersymmetry is only
one of many possible scenarios providing corrections to $a_\mu$ at the weak
scale.  Theoretical prejudice tends to favor supersymmetry, but other
possibilities exist, summarized in Ref.~\cite{SM}.

\section{supersymmetric model}

In the Minimal Supersymmetric Standard Model (MSSM) the lightest of the
superpartners (LSP) is often the lightest neutralino.  These four states are
superpositions of the superpartners of the neutral gauge and Higgs bosons,
\begin{equation}
\tilde{\chi}^0_1 =
N_{11} \tilde{B} + N_{12} \tilde{W}^3 +
N_{13} \tilde{H}^0_1 + N_{14} \tilde{H}^0_2.
\end{equation}
With R-parity conserved, this lightest superpartner is stable.  For significant
regions of the MSSM parameter space, the relic density of the stable neutralino
is of the order $\Omega_\chi h^2\sim0.1$, thus constituting an important (and
perhaps exclusive) part of the cold dark matter (for a review see
Ref.~\cite{jkg96}).  Note that $\Omega_{\chi}$ is the neutralino density in
units of the critical density and $h$ is the present Hubble constant in units
of $100$ km s$^{-1}$ Mpc$^{-1}$.  Large scale structure observations favor $h =
0.7\pm 0.1$ and a matter density $\Omega_{M} = 0.3 \pm 0.1$, of which baryons
contribute a small amount $\Omega_bh^2=0.02\pm 0.001$ \cite{cosmparams}.  CMB
anisotropy measurements are consistent with this (summarized in e.g.\
\cite{cmbdata}), favoring $\Omega_Mh^2=0.15\pm0.05$.  We take the range
$0.05\le\Omega_\chi h^2\le0.25$ as the cosmologically interesting region,
basically a 2$\sigma$ constraint.  Models where neutralinos are not the only
component of dark matter are also allowed, so we separately consider
arbitrarily small $\Omega_\chi h^2<0.25$.  Even with very small relic
densities, such models may be observable in astrophysical contexts
\cite{lowOmega}.

We have explored a phenomenological variation of the MSSM with seven free
parameters: the higgsino mass parameter $\mu$, the gaugino mass parameter
$M_{2}$, the ratio of the Higgs vacuum expectation values $\tan \beta$, the
mass of the $CP$--odd Higgs boson $m_{A}$, the scalar mass parameter $m_{0}$
and the trilinear soft SUSY--breaking parameters $A_{b}$ and $A_{t}$ for third
generation squarks.  All of our parameters are fixed at the electroweak scale.
Our framework is more general than the supergravity framework, in that we do
not impose GUT unification of the scalar masses and trilinear couplings.  In
contrast to supergravity, this allows a highly pure higgsino LSP and its
consequences, namely a SUSY spectrum that can be significantly more massive.
For simplicity, we do apply the supergravity constraint on gaugino mass
unification, though the relaxation of this constraint would not significantly
alter our results.  As is typical, we assume R-parity conservation, stabilizing
the lightest superpartner.  (These models are described in more detail in
Refs.~\cite{bg,coann,jephd}.)

\begin{table}
\begin{ruledtabular}
\begin{tabular}{rrrrrrrr}
Parameter & $\mu$ & $M_{2}$ & $\tan \beta$ & $m_{A}$ & $m_{0}$ &
$A_{b}/m_{0}$ & $A_{t}/m_{0}$ \\
Unit & TeV & TeV & 1 & TeV & TeV & 1 & 1 \\ \hline Min & -50 &
-50 & 1.0 & 0        & 0.1 & -3 & -3 \\
Max & 50 & 50 & 60.0 & 10 & 30 & 3 & 3 \\
\end{tabular}
\end{ruledtabular}
\caption{The ranges of parameter values used in the MSSM scans of
Refs.~\protect\cite{bg,coann,bub,neutrate,other_db}.  We use approximately
80,000 models not excluded by accelerator constraints or the cosmological relic
density bound ($\Omega_\chi h^2<0.25$) before the $a_\mu$ measurement.
Approximately 25,000 of these lie in the cosmologically interesting
region ($0.05<\Omega_\chi h^2<0.25$).}
\label{tab:scans}
\end{table}

To investigate the MSSM parameter space, we have used the database of MSSM
models built in Refs.~\cite{bg,coann,bub,neutrate,other_db}.  Furthermore, for
this work we have made special scans emphasizing large positive supersymmetric
corrections to $a_\mu$, $\dams$.  The overall ranges of the seven MSSM
parameters are given in Table~\ref{tab:scans}.  The database includes one--loop
corrections for the neutralino and chargino masses as given in
Ref.~\cite{neuloop}, and leading log two--loop radiative corrections for the
Higgs boson masses as given in Ref.~\cite{feynhiggs}.  Supersymmetric
contributions to the precision quantities $a_\mu$ and the $b\to s \gamma$
branching ratio are also included.  The database contains the
neutralino--nucleon cross sections and expected detection rates for a variety
of neutralino dark matter searches.

Crucial for studies of dark matter, the database includes the cosmological
relic density of neutralinos $\Omega_{\chi} h^2$, based on calculations in
Refs.~\cite{coann,GondoloGelmini} considering resonant annihilations, threshold
effects, finite widths of unstable particles, all two--body tree--level
annihilation channels of neutralinos, and coannihilation processes between all
neutralinos and charginos.  We are in the process of including a complete
treatment of sfermion coannihilations \cite{sfermioncoann}.

Recent accelerator constraints are applied to each model in the database.  Most
important are the LEP bounds \cite{pdg2000} on the lightest chargino mass
(chargino and neutralino masses are tightly linked)
\begin{equation}
m_{\chi_{1}^{+}} > \left\{
\begin{array}{lcl}
88.4 {\rm ~GeV} & \quad , \quad & | m_{\chi_{1}^{+}} - m_{\chi^{0}_{1}} | > 3
{\rm ~GeV} \\ 67.7 {\rm ~GeV} & \quad , \quad & {\rm otherwise,} \end{array}
\right.
\end{equation}
and on the lightest Higgs boson mass $m_{h}$ (which ranges from 91.5--112 GeV
depending on $\tan\beta$) and the $b \to s \gamma$ branching ratio \cite{cleo}
(DarkSUSY currently only implements the leading-order calculation
\cite{DarkSUSY}).

\section{muon anomalous magnetic moment}

Supersymmetric corrections to $a_\mu$ are surprisingly large, enhanced relative
to typical weak--scale contributions by the parameter $\tan\beta$
\cite{a_mu_old,moroi,SM}.  This fact makes these precision measurements
enticing approaches for searching for supersymmetry.  Typically, the
supersymmetric corrections are given by
\begin{equation}
\dams\sim 14\times10^{-10}\left(\frac{M_{\rm SUSY}}{100\;\rm GeV}\right)^{-2}
\tan\beta,
\end{equation}
where $M_{\rm SUSY}$ is the typical mass of superpartners.

The new results of Brookhaven AGS experiment E821 \cite{newdata} for the
anomalous magnetic moment of the muon, $a_\mu=(g-2)/2$, compared with the
predicted Standard Model value are
\begin{eqnarray}
a_\mu({\rm exp})&=&11\;659\;204(8)\times10^{-10}, \\
a_\mu({\rm SM})&=&11\;659\;175(7)\times 10^{-10}, \\
\Delta a_\mu&=&29(11)\times 10^{-10}.
\label{eq:disc}
\end{eqnarray}
This indicates a disagreement with the Standard Model at a $2.6\sigma$
confidence level.  However, as mentioned in the introduction, there exist newer
and conflicting evaluations of the Standard Model contribution which increase
or decrease the discrepancy significantly.  We will focus on the slightly older
evaluation, but we will qualitatively discuss the effects of applying either of
the new evaluations.  To investigate the implications for the supersymmetric
parameter space, we will assume that supersymmetry is the only source of
corrections to $a_\mu$ outside of the Standard Model. Considering a 95\%
$(2\sigma)$ confidence region for the supersymmetric contribution, we accept
the following range of $\dams$
\begin{equation}
7\times10^{-10}\le\dams\le51\times10^{-10}.
\end{equation}
We have used the full calculation in Ref.~\cite{moroi} to compute $\dams$ for
the models in the database.

\begin{figure*}[!ht]
\centerline{
\epsfig{file=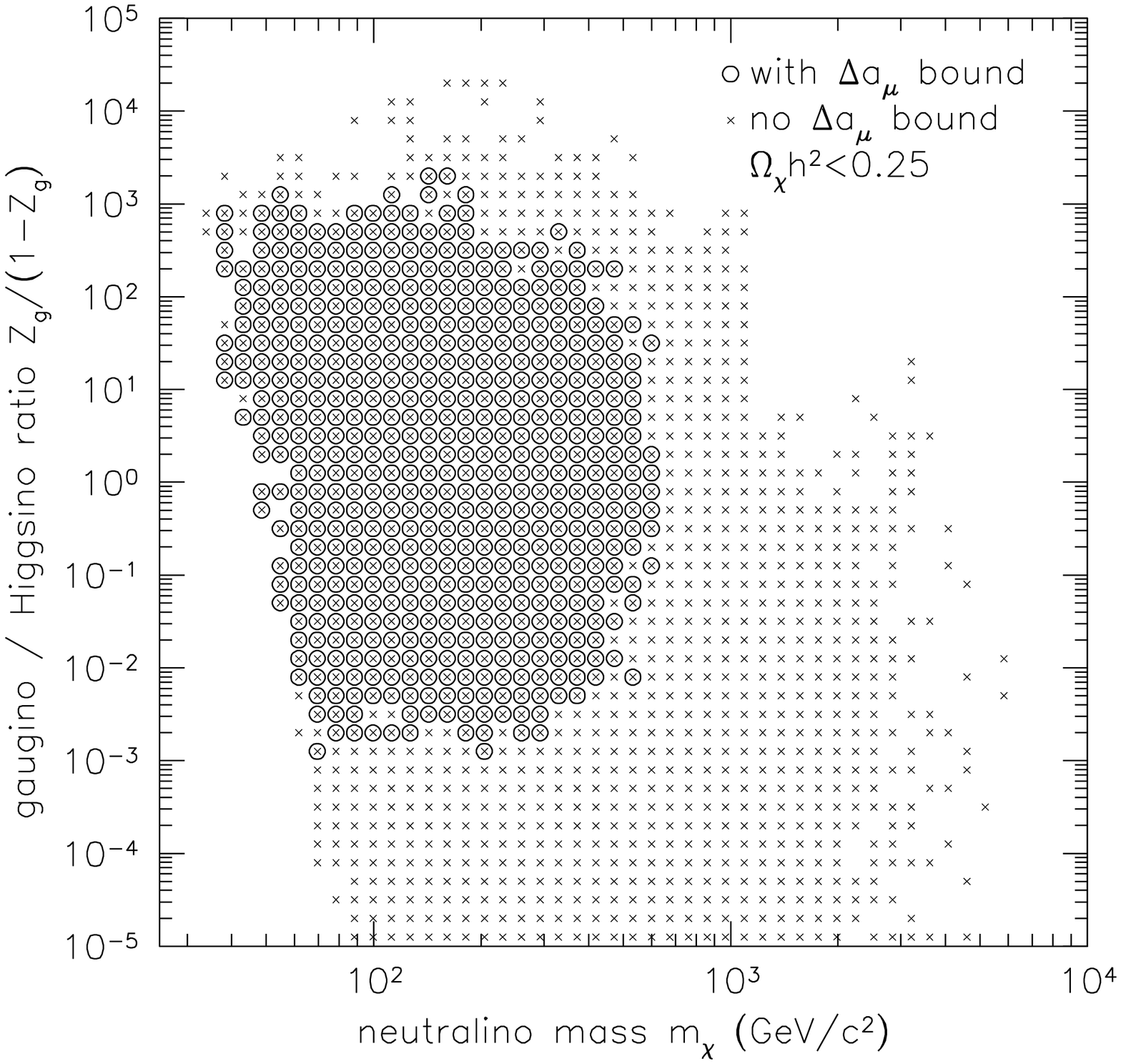,width=0.415\textwidth}
\epsfig{file=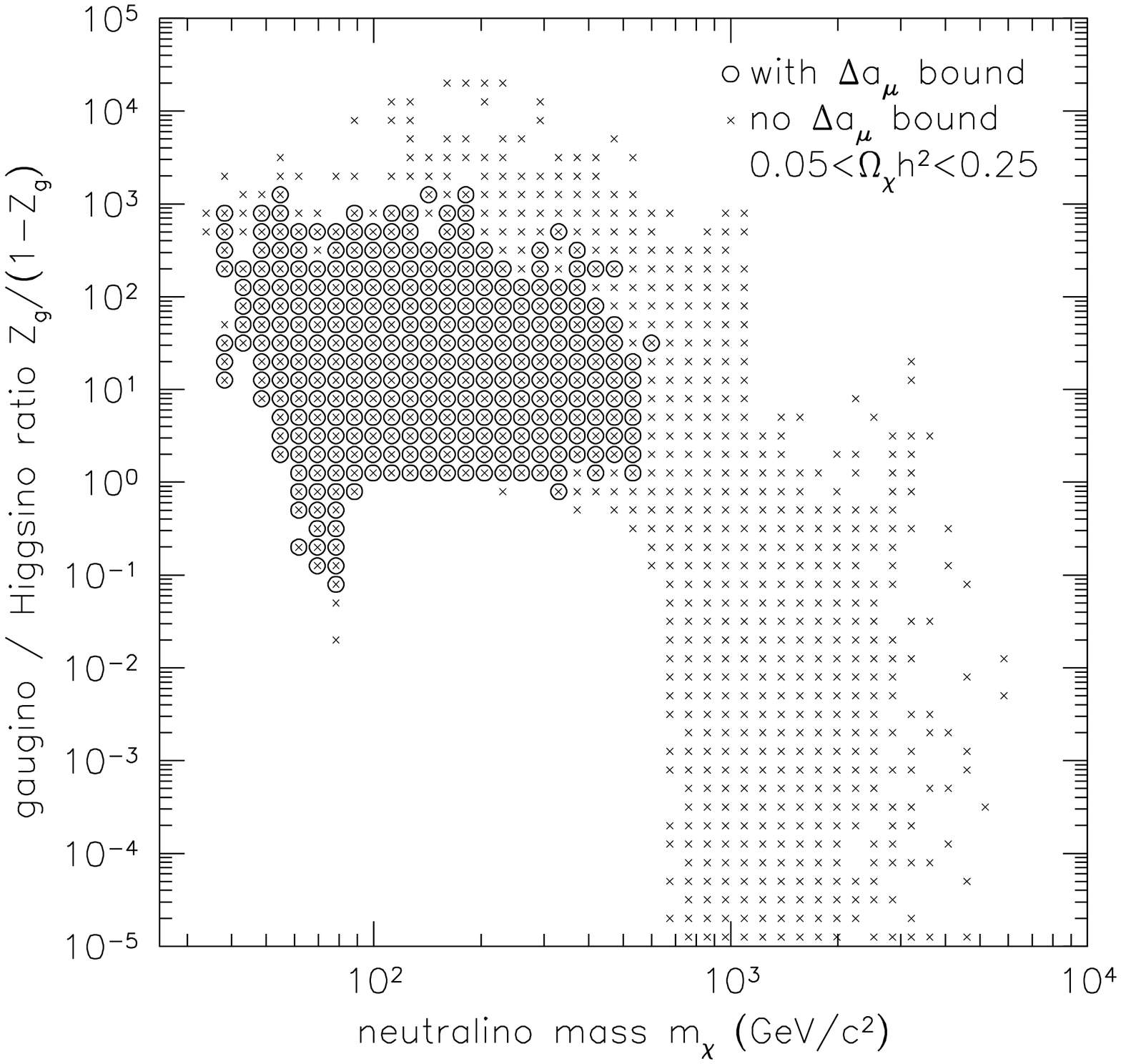,width=0.415\textwidth}
}
\caption{Gaugino/higgsino fraction versus mass for the lightest neutralino.  In
the left panel, we plot all models not excluded by cosmological arguments.  In
the right panel, only models with an interesting relic density are plotted.
Crosses indicate models allowed before applying a constraint on $\dams$, and
crossed circles indicate models allowed after imposing the $\dams$ bound.}
\label{fig:mxzg}
\end{figure*}

The astrophysical phenomenology of neutralinos depends strongly on
the ratio of gaugino and higgsino fractions, defined as
\begin{equation}
\frac{Z_g}{1-Z_g} = \frac{ |N_{11}|^2 + |N_{12}|^2}{|N_{13}|^2 + |N_{14}|^2}.
\end{equation}
We plot this ratio against the neutralino mass for each model in the database
in Fig.~\ref{fig:mxzg}.  For clarity, the models have been binned along both
axes.  In the left panel, we only require that $\Omega_\chi h^2<0.25$, and on
the right we apply the more stringent constraint that the neutralino could make
up all of the dark matter, $\Omega_\chi h^2=0.15\pm0.1$.  Models allowed before
the new $\dams$ constraint are plotted as crosses, and models respecting the
new $\dams$ constraint are plotted as crossed circles.

As has been discussed at length previously \cite{bg2001,otherpapers}, a $\dams$
bound that excludes zero from the positive side gives an upper limit on the
mass of the neutralino, in this case 650 GeV.  This is a large improvement over
the cosmological bound based on the neutralino relic density not being too
large, an upper limit of 7 TeV \cite{coann}.  However, if the Standard Model
value is included in the allowed region by e.g.\ considering a $3\sigma$
confidence interval or a revised Standard Model calculation, there is no bound
on neutralino mass.  If the more favorable evaluation is considered, the mass
bound becomes 450 GeV.

If the neutralino has a large enough relic density to make up all of the cold
dark matter, another effect appears, namely that the neutralino can not be very
purely higgsino-like in composition, requiring at least a 5\% mixture (in
quadrature) of gaugino states.  Even without requiring neutralino dark matter,
higgsino-like neutralinos are disfavored, with a maximum purity of 99.9\%
($Z_g=0.001)$.

\begin{figure*}[!ht]
\centerline{
\epsfig{file=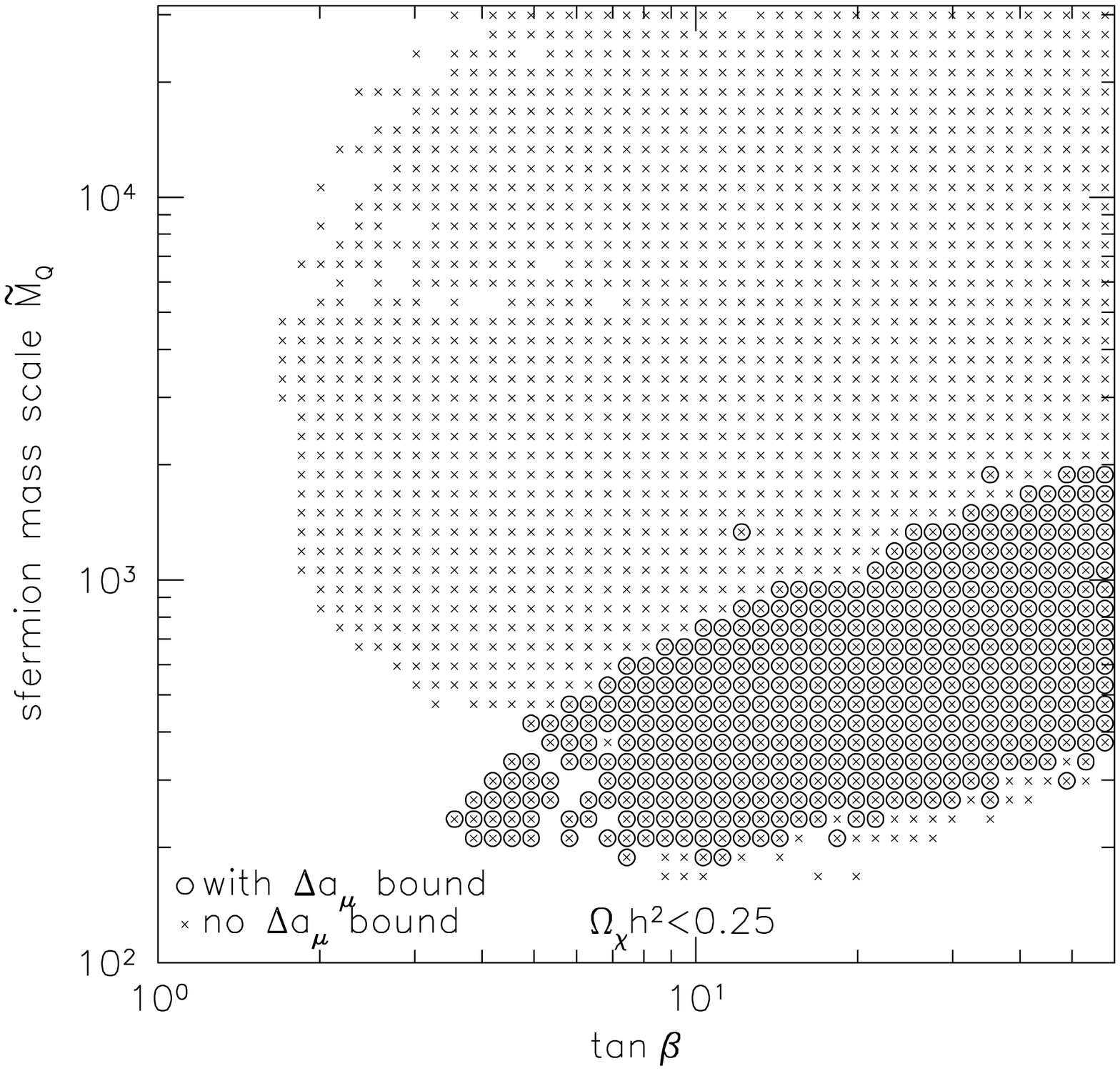,width=0.415\textwidth}
\epsfig{file=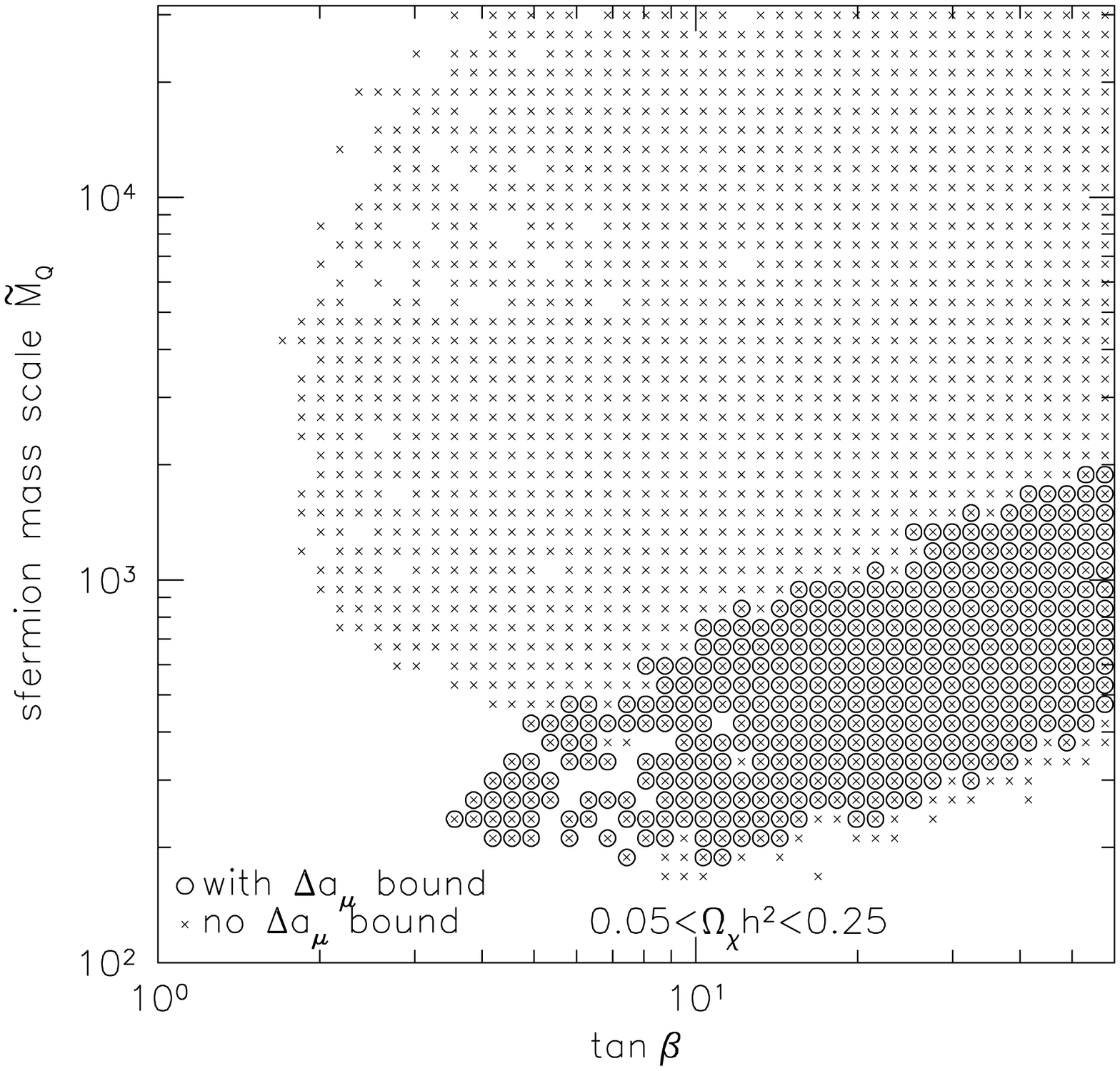,width=0.415\textwidth}
}
\caption{Sfermion mass scale versus $\tan\beta$.  As in Fig.~\ref{fig:mxzg}, in
the left panel, we plot all models not excluded by cosmological arguments and
in the right panel we plot only models with an interesting relic density.
Crosses indicate models allowed before applying a constraint on $\dams$, and
crossed circles indicate models allowed after imposing the $\dams$ bound.  The
small ``inlet'' at $\tan\beta\sim4$ and $m_0\sim400$ GeV is due to the
constraint on the Higgs boson mass.  It is clear that the relic density cut has
little effect on the allowed region.}
\label{fig:tanbm0}
\end{figure*}

Orthogonal to the neutralino mass and composition but equally important to the
value of $\dams$ are the parameters $\tan\beta$ (the ratio of vacuum
expectation values) and $m_0$ (the scalar mass parameter).  In
Fig.~\ref{fig:tanbm0} we plot these parameters for the database of models,
again indicating the effects of the constraint on $\dams$.  The constraint
forces the scalar mass parameter to be small, but the upper bound increases
with increasing $\tan\beta$.

\section{astrophysical dark matter searches}

There is a large community pursuing the goal of detecting dark matter
particles, neutralinos especially, in various astrophysical contexts.  The
possibilities can be broken up along the lines of ``direct'' and ``indirect''
detection.  Direct detection means detecting the rare scatterings of
neutralinos in our galactic halo with nuclei in a sensitive low background
apparatus.  Indirect detection means detecting the products of rare
annihilations of galactic neutralinos, such as antiprotons, antideuterons,
positrons, gamma rays, and neutrinos.

\begin{figure*}[!ht]
\centerline{
\epsfig{file=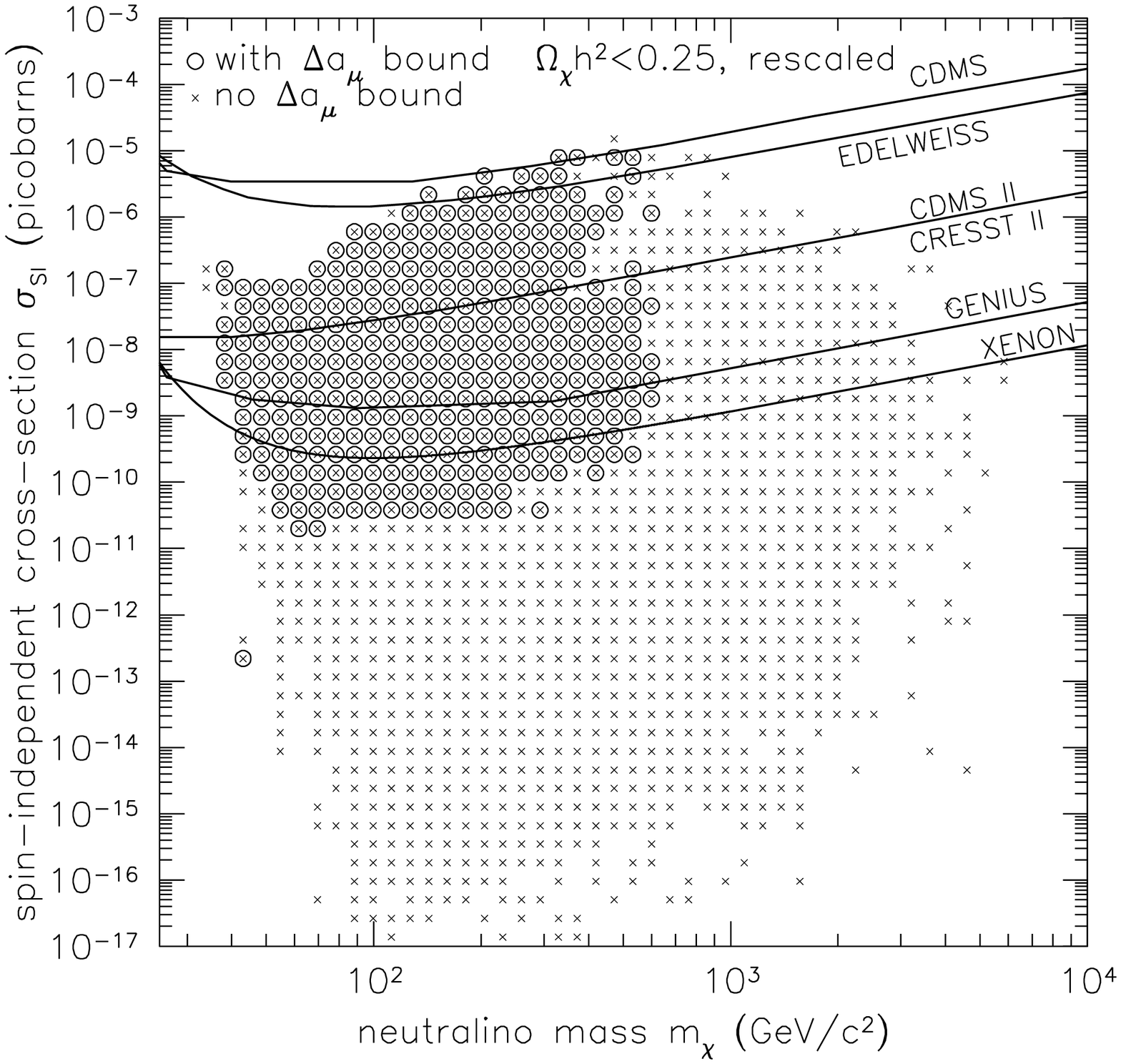,width=0.415\textwidth}
\epsfig{file=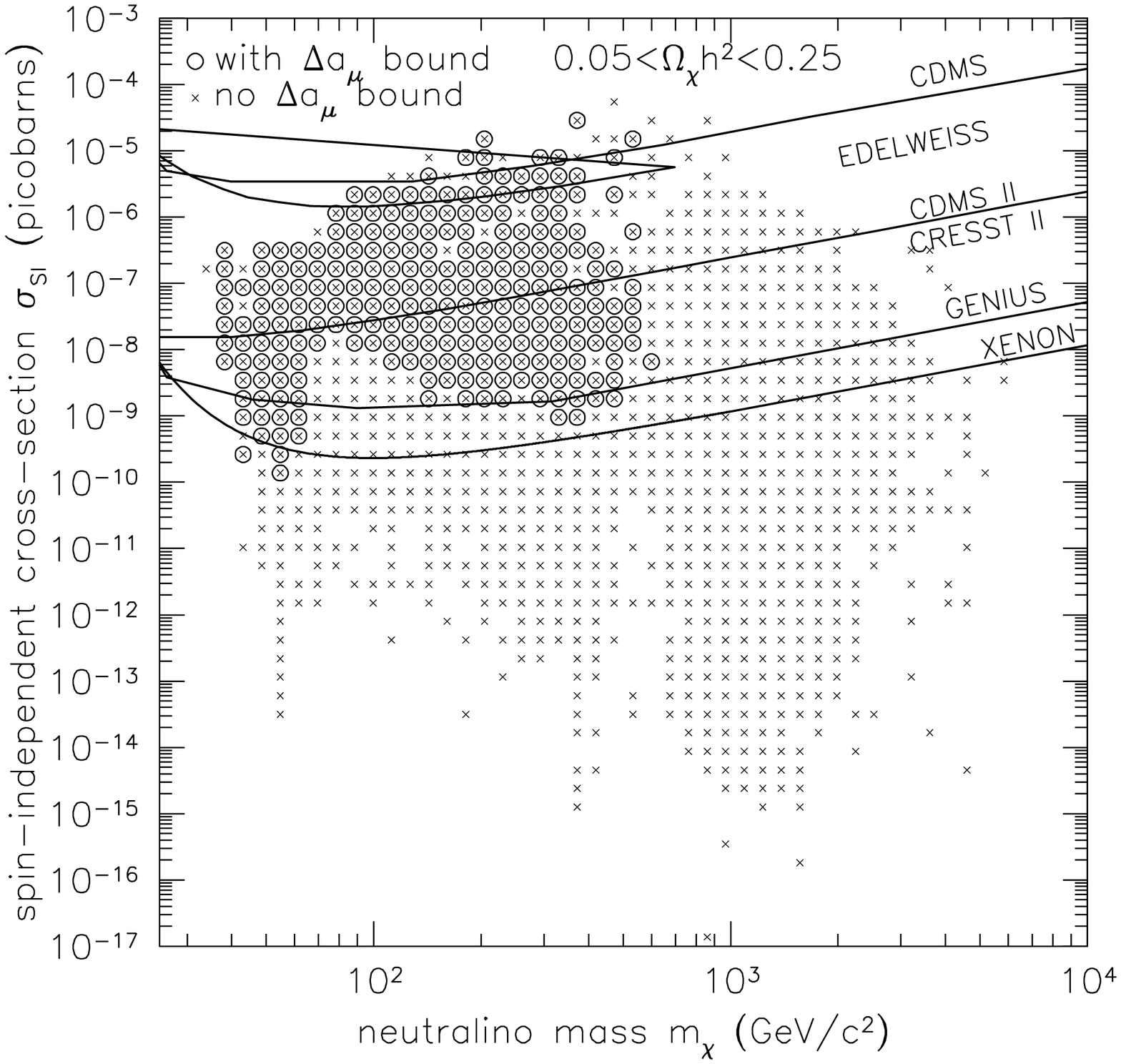,width=0.415\textwidth}
}
\caption{Neutralino--nucleon elastic scattering cross section versus neutralino
mass.  In the left panel, we have only applied the upper constraint on relic
density, and rescaled the effective cross section to account for a low galactic
density of low relic density neutralinos.  In the right panel we plot only
those models that could account for all of the dark matter, and we do not
perform a rescaling.  The inlet at 100 GeV and $10^{-9}$ pb is due to the lower
limit $\Omega_\chi h^2>0.05$.  Crosses indicate models allowed before applying
a constraint on $\dams$, and crossed circles indicate models allowed after
imposing the $\dams$ bound.}
\label{fig:directdetect}
\end{figure*}

Perhaps the most promising of the astrophysical neutralino searches is the
direct detection program.  Experiments such as CDMS \cite{cdms}, DAMA
\cite{dama}, and EDELWEISS \cite{edelweiss} have pushed exclusion limits down
to cross sections as small as $10^{-6}$ pb.  As has been noted before, the
neutralino--nucleon elastic scattering cross section exhibits a significant
correlation with $\dams$ \cite{drees,bg2001}, thus a large positive $\dams$ is
exciting for direct searches.  Direct detection is promising even in the case
where the neutralinos have a small relic density and thus are only a small
component of the dark matter.  In this case we perform a conservative rescaling
of the galactic neutralino density as
\begin{equation}
\rho_\chi\rightarrow\rho_\chi\left(\frac{\Omega_\chi h^2}{0.25}\right),
\label{eq:rescale}
\end{equation}
where $\Omega_\chi h^2=0.25$ is the current upper limit on the relic density.
In the left panel of Fig.~\ref{fig:directdetect}, we plot the spin-independent
neutralino-proton scattering cross section, rescaled according to
Eq.~\ref{eq:rescale}.  The constraint due to $\dams$ is intriguing, as it
bounds the rescaled cross section at around $10^{-11}$ pb.  In the right panel
of Fig.~\ref{fig:directdetect}, we perform no rescaling, and only consider
models with cosmologically interesting relic densities.  Here the minimum cross
section is around $10^{-10}$ pb.  The inlet at 100 GeV and $10^{-9}$ pb is due
to the lower limit $\Omega_\chi h^2>0.05$.  These bounds indicate that there is
considerable hope for the next generation of experiments, such as CDMS II and
CRESST II \cite{cresst}.  The latter bound is perhaps reachable by future
experiments with one ton target masses such as GENIUS \cite{genius}, CryoArray
\cite{cryoarray}, and XENON \cite{xenon}.  Finally, it is important to note
that in the case where the significance of the $a_\mu$ discrepancy is reduced,
the lower bound on the cross section disappears.  However, there is {\em not}
an upper bound on the cross section.  Large cross sections are still possible
with $\dams$ consistent with zero.

\begin{figure*}[!ht]
\centerline{
\epsfig{file=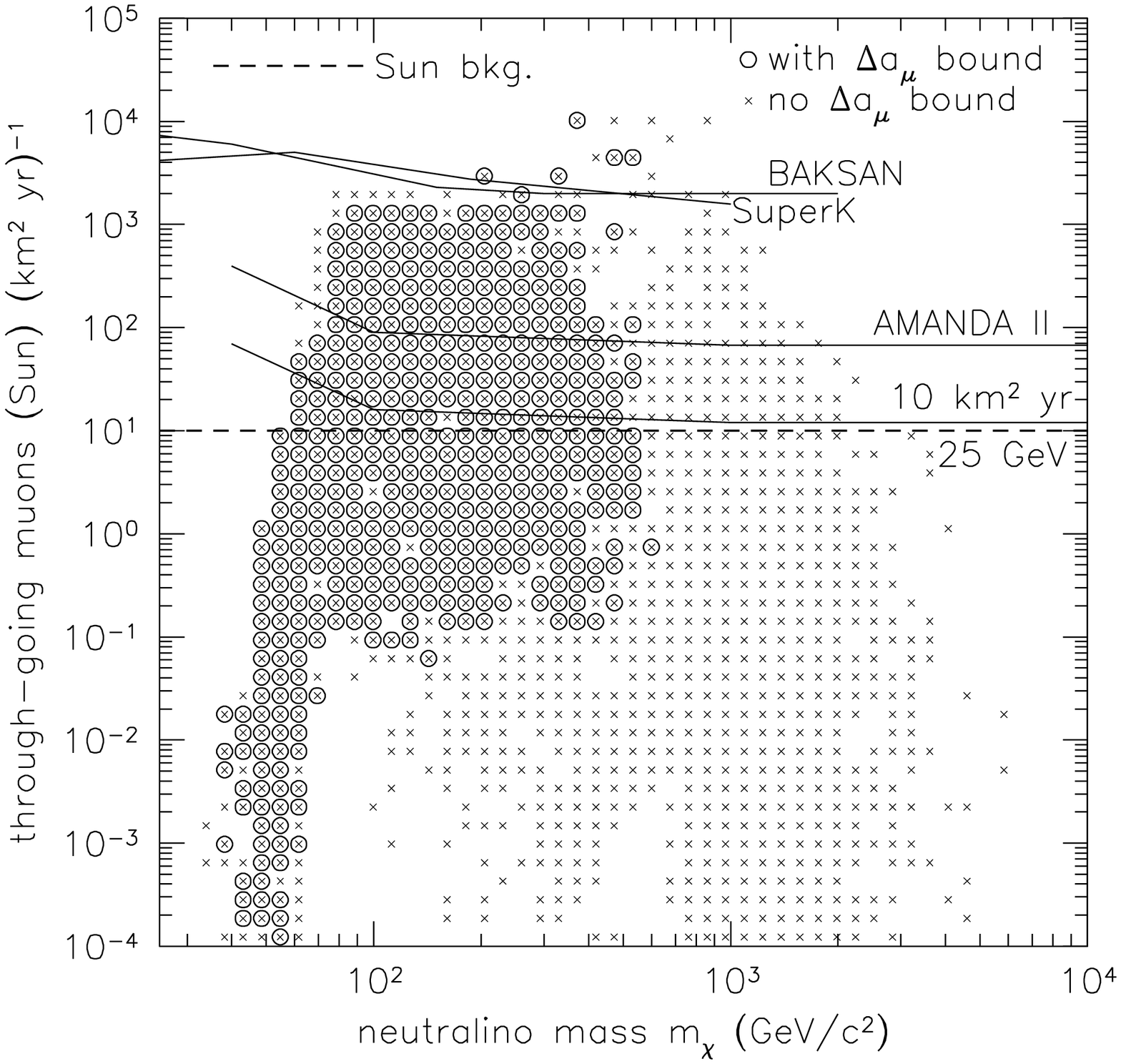,width=0.415\textwidth}
\epsfig{file=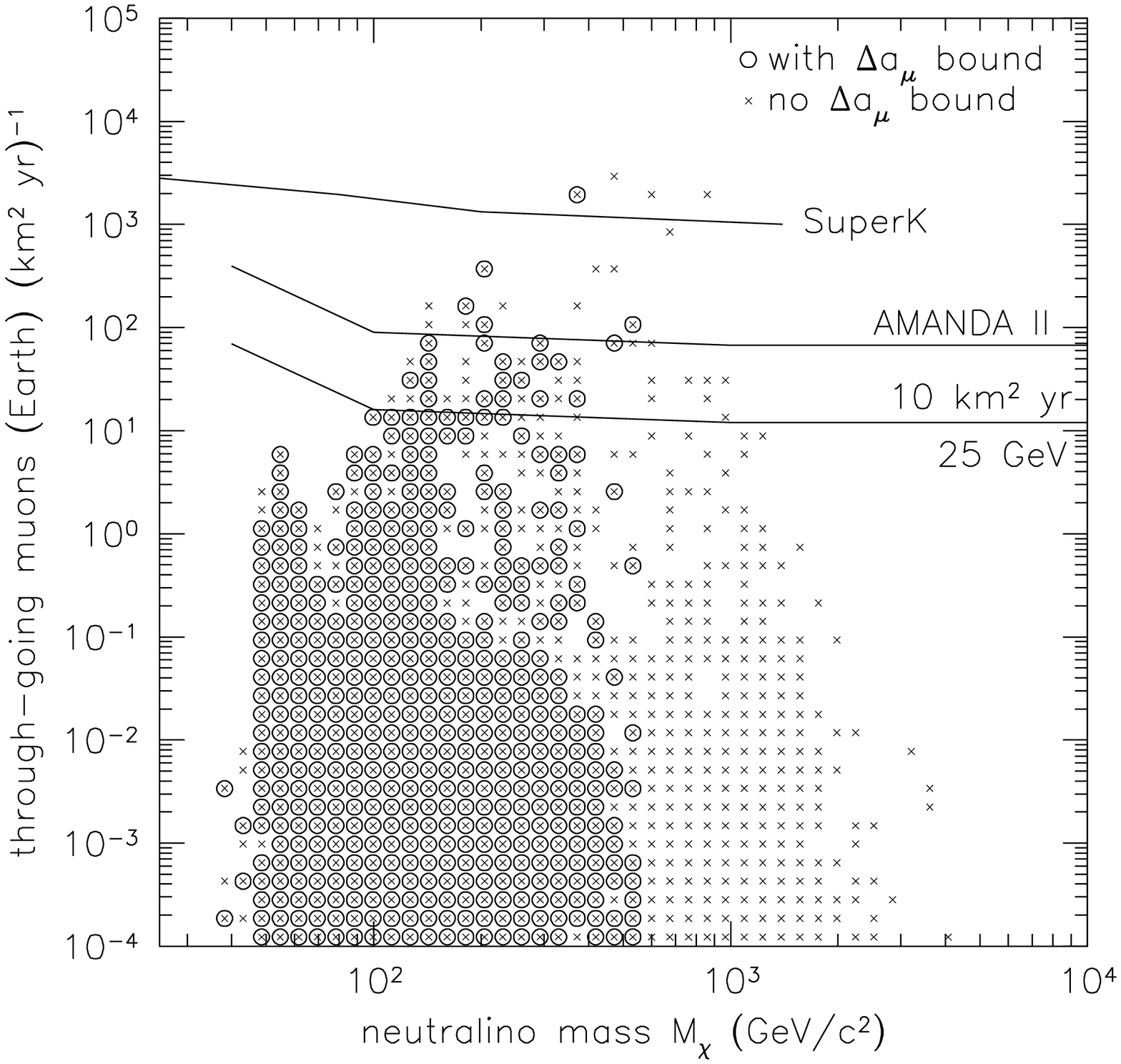,width=0.415\textwidth}
}
\centerline{
\epsfig{file=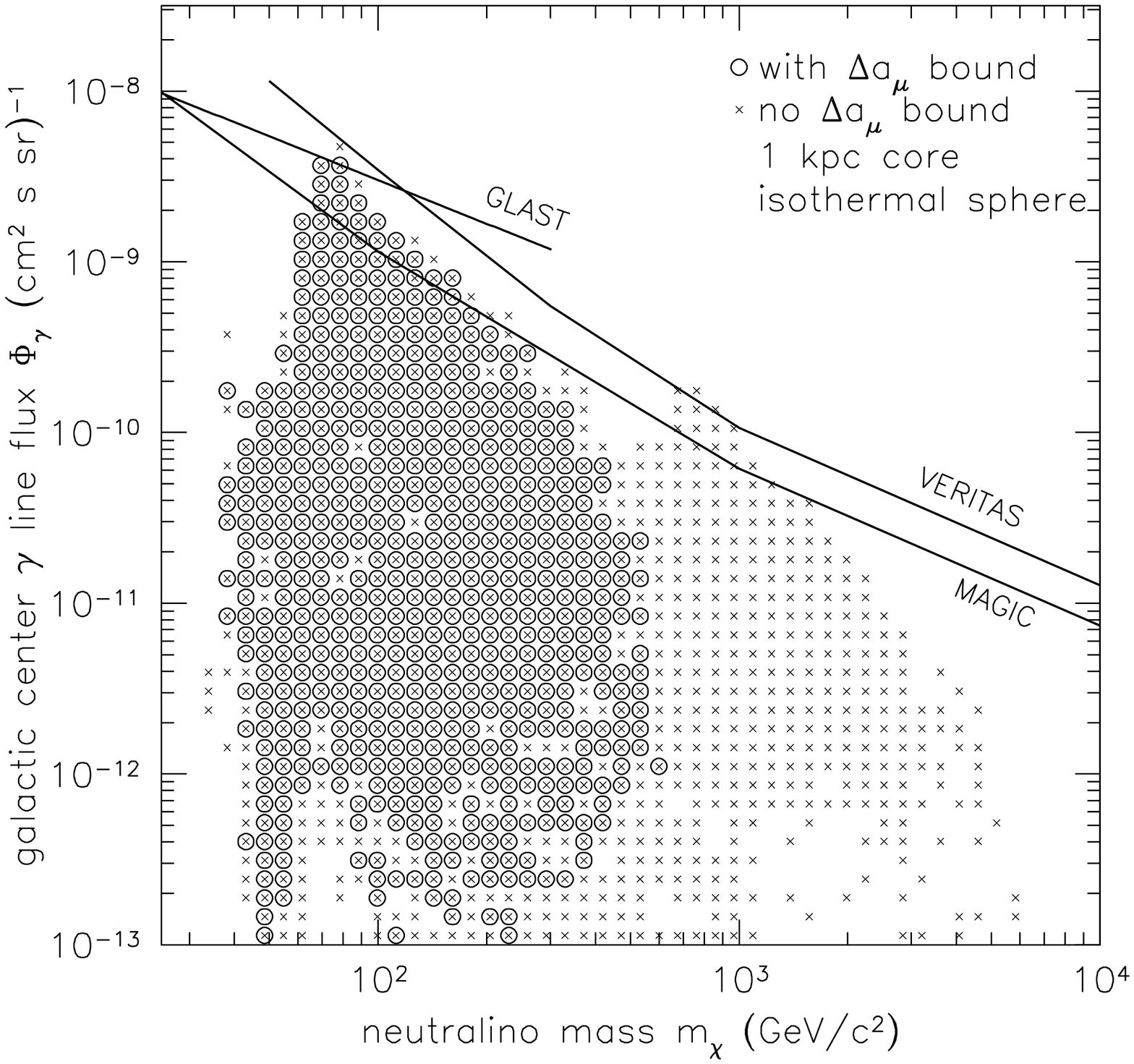,width=0.415\textwidth}
\epsfig{file=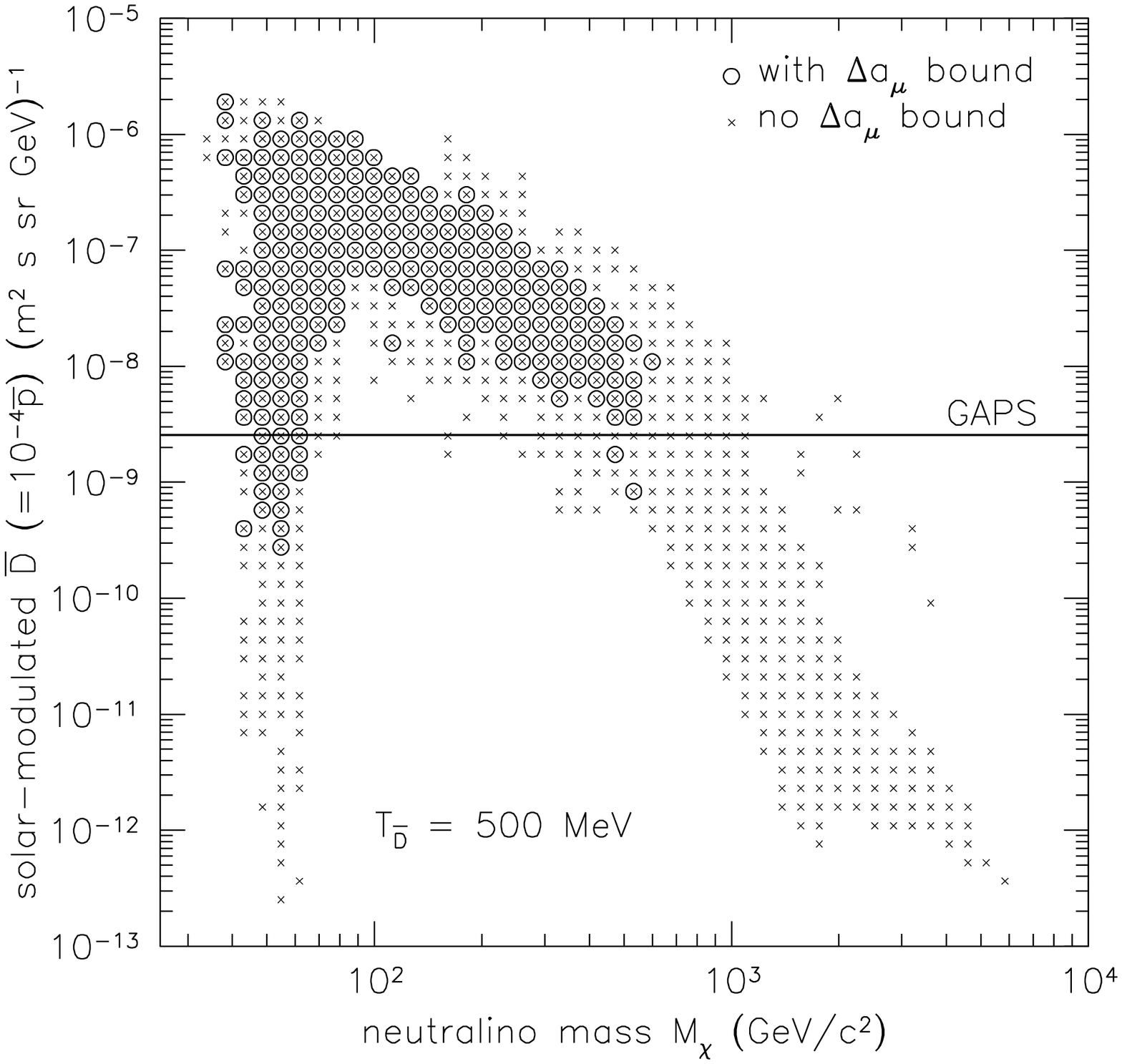,width=0.415\textwidth}
}
\caption{Indirect detection of neutralinos.  In all plots, small crosses
indicate cosmologically interesting models $(0.05<\Omega_\chi h^2<0.25)$, and
crossed circles indicate such models that pass the $\dams$ cut.  In the top
left we plot the rate of through-going muons in a neutrino telescope for the
annihilations in the Sun, with the BAKSAN and SuperKamiokande bounds, and the
reach of a km$^2$ telescope.  In the top right, we plot a similar rate for
neutrinos from the center of the Earth.  In the bottom left we plot the flux in
the gamma ray lines from the galactic center, assuming a 1 kpc core isothermal
sphere halo.  The future reach of the GLAST, VERITAS, and MAGIC experiments is
included.  In the bottom right we plot the flux of antideuterons at a kinetic
energy of 500 MeV, together with the future sensitivity of the GAPS detector.}
\label{fig:indirect}
\end{figure*}

Neutrino telescopes such as at Lake Baikal \cite{baikal}, Super-Kamiokande
\cite{superk}, in the Mediterranean \cite{antares}, and the south pole
\cite{amanda} are a promising technique for indirect detection.  Neutralinos in
the galactic halo scatter into orbits around the Earth or Sun, and can then
rapidly sink to the cores of these bodies by additional scatterings, resulting
in a large density enhancement.  This can produce a detectable annihilation
signal in neutrinos at high (GeV) energies.  It is the capture rate that
governs the neutrino flux, and is strongly correlated with the
neutralino--nucleon cross section.  This places a lower bound on the detection
rate, though at small neutralino mass there are threshold effects that remove
it \cite{neutrate}.  To illustrate, we plot the rate of neutrino-induced
through-going muons from the Sun along with the unsubtractable background (from
cosmic rays incident on the Sun's surface) in the top left panel of
Fig.~\ref{fig:indirect}.  It is clear that the $\dams$ bound cuts away much of
the undetectable parameter space, but not all of it.  In the top right panel we
repeat the calculation for neutrinos from the center of the Earth, where the
prospects are much less promising.

In addition, neutralinos can annihilate in the galactic halo.  The relevant
rates are quite small, but the enormous mass of the halo compensates, and the
annihilation products may be detectable.  Gamma rays propagate essentially
freely, thus the expected rate is largest towards the galactic center where the
dark matter density is largest.  Charged particles are trapped by the galactic
magnetic field and effectively diffuse, so these annihilation products would
originate more nearby.

The detection of the gamma ray lines from direct annihilations either to two
photons, or to a photon and a $Z$ boson would be a gold--plated signature of
neutralinos in the galactic halo \cite{bub}.  Gamma ray experiments such as the
atmospheric \v Cerenkov telescopes (ACTs) VERITAS \cite{veritas} and MAGIC
\cite{magic}, and the GLAST \cite{glast} satellite hope to detect these lines.
Assuming that the galactic halo is an isothermal sphere with a 1 kpc core, we
plot the reach of these experiments in the bottom left panel of
Fig.~\ref{fig:indirect}.  Note that with this assumption the emission
enhancement from around the black hole at the galactic center is insignificant
\cite{blackhole}, so we neglect it.  We notice that applying the $\dams$ bound
has little effect on the prospects for these experiments.  It appears that
the detection of the gamma ray lines is quite difficult.  Other assumptions,
including clumping of the dark matter or a lack of a central core lead to
significantly higher predictions \cite{bub}.

The intensity of cosmic ray positrons \cite{positrons} from neutralino
annihilation is not much affected by the constraint on $\dams$.

The final possibility we mention is that antideuterons may be an interesting
annihilation product to search for \cite{dbar}.  The background from mundane
cosmic ray processes should be relatively smaller (a smaller fraction of the
annihilation signal) at low energies than for antiprotons.  A signal may be
detectable in experiments such as AMS \cite{ams} and GAPS \cite{gaps}, as seen
in the bottom right of Fig.~\ref{fig:indirect}.  It is interesting that the
$\dams$ constraint eliminates the models with the lowest rates, and furthermore
that the whole parameter space is covered for neutralino masses between 100 GeV
and 500 GeV.  Separating a signal from the background with antideuterons may be
difficult however.

\section{conclusions}
In this paper we have discussed the recent confirmation of a discrepancy with
the Standard Model of the anomalous magnetic moment of the muon \cite{newdata},
updating and expanding the results of Ref.~\cite{bg2001}.  Assuming that
supersymmetry is responsible for the discrepancy, we have investigated the
consequences for astrophysical dark matter searches.  We have confirmed that
the constraint significantly improves the prospects for direct detection
experiments trying to measure the rare scatterings of galactic neutralinos.
Neutrino telescopes are also helped by this result.  The prospects for the
detection of gamma ray lines from neutralino annihilations at the galactic
center are not much affected.  The prospects for detecting cosmic ray
antideuterons as neutralino annihilation products are also significantly
improved.  In all cases, if the discrepancy disappears, there remain
supersymmetric models with detectable rates for all of these experiments.

\begin{acknowledgments}
E.B.\ thanks W.~Marciano and W.~Morse for useful conversations.
\end{acknowledgments}


\end{document}